%% file: wpaper.tex
\documentclass[sigconf]{acmart}

\usepackage{booktabs} 
\newcommand{\burl}[1]{{\color{blue}\url{#1}}}

\newcommand{\btt}{\usefont{OT1}{lmtt}{b}{n}}
\setcopyright{rightsretained}

\acmConference[TempWeb2018]{Temporal Web Analytics 2018}{April 2018}{Lyon, France} 
\acmYear{2018}
\copyrightyear{2018}

\begin{document}
\title{Can Common Crawl reliably track persistent identifier (PID) use over time?}
\renewcommand{\shorttitle}{PIDs in Common Crawl}

\author{Henry S. Thompson}
\orcid{0001-5490-1347}
\affiliation{%
  \institution{University of Edinburgh}
  \streetaddress{Informatics Forum 4.22\
10 Crichton Street}
  \city{Edinburgh} 
  \postcode{43017-6221}
  \country{United Kingdom}
}
\email{ht@inf.ed.ac.uk}

\author{TONG Jian}
\authornote{SURNAME forename}
\affiliation{%
  \institution{University of Edinburgh}
  \streetaddress{Informatics Forum\
10 Crichton Street}
  \city{Edinburgh} 
  \postcode{43017-6221}
  \country{United Kingdom}
}
\email{s1615354@sms.ed.ac.uk}

\begin{abstract}
  We report here on the results of two studies using two and four
  monthly web crawls respectively from the Common Crawl (CC)
  initiative between 2014 and 2017, whose initial goal was to provide
  empirical evidence for the changing patterns of use of so-called
  persistent identifiers.  This paper focusses on the tooling needed
  for dealing with CC data, and the problems we found with it.  The
  first study is based on over $10^{12}$ URIs from over $5 x 10^9$
  pages crawled in April 2014 and April 2017, the second study adds a
  further $3 x 10^9$ pages from the April 2015 and April 2016 crawls.
  We conclude with suggestions on specific actions needed to enable
  studies based on CC to give reliable longitudinal information.
\end{abstract}

%
%

\keywords{temporal web analytics,
persistent identifier,
Common Crawl,
Uniform Resource Identifier,
longitudinal web crawl analysis,
Digital Object Identifier}

\maketitle

\input{wpaperbody}

\vspace{1cm}
\bibliographystyle{ACM-Reference-Format}
\bibliography{wpaper} 

\end{document}

%% file: wpaperbody.tex
\section{Introduction}

The history of efforts to meet the demand for so-called `persistent
identifiers' (PIDs) for use on the Web is complicated, with many
alternative offerings and much debate about the meaning of persistence
and how to go about ensuring it.  We take no position in that debate
here, beyond the observation that the demand for PIDs shows no signs
of abating, and that there has been a more-or-less general
acknowledgement over the last 5--10 years that to be successful in the
context of the Web a PID scheme must define and support a mapping
from PIDs in the scheme to `actionable' identifiers. In practice this
has meant specifying a purely syntactic procedure for converting a PID
into an {\tt http(s):} URI using a domain owned and operated by the
proprietors of the scheme.  An HTTP request for such `actionable' URIs
will typically result in a redirection to the then-current location of
the identified resource.

The Digital Object Identifier scheme \cite{DOI}, managed by the
International DOI Foundation (IDF) \cite{IDF}, was an early adopter of this
approach, and DOIs are now in widespread use, particularly in
scientific journals, where their use is actually mandated by a number
of major publishers.  The mapping for DOIs to actionable {\tt https:} URIs is
simple: For example a DOI for a journal article written in the form of
a URI such as \url{doi:...} is
mapped (client-side) to \burl{https://doi.org/...}\footnote{{\tt doi:} is not (yet) a registered
  URI scheme, but often used as if it were one}. In response to an HTTP
request for that URI, the server at {\tt doi.org} (operated on behalf of IDF
by the Corporation for National Research Initiatives (CNRI) \cite{CNRI})
will respond with a redirect to the appropriate {\tt http(s):} URI from the
actual publisher of the article.  We call the three forms involved the
'original' (e.g. {\tt doi:} or {\tt info:hdl}), the `actionable'
(e.g. {\tt https://doi.org/...} and variants
thereof or {\tt http://hdl.handle.net/...}) and the `locating'.  Note that {\em none} of these is
strictly speaking a PID as such: that's what comes {\em after} the
{\tt doi:} or {\tt https://hdl:handle.net/}.

The success of this approach has overcome a significant barrier to
the adoption of PIDs in general:  to date there has been no
significant move towards support for any of them {\em as URIs} in web
browsers or PDF viewers.  That is, if you try to use \burl{doi://10.1000/182}
or \burl{info:hdl/20.1000/100}
as a link (for example, as the value of the
{\btt href} attribute of an HTML {\btt A} element), it will not work.  But you
{\em can} use them as the link text of an {\btt A} element, and put the actionable form
(\burl{https://doi.org/10.1000/182} and
\burl{http://hdl.handle.net/20.1000/100} respectively) in the {\btt href} attribute, and {\em that} will work just fine.

That's the good news.  The less good news is that the use of
redirection from the actionable form to the locating form means that
when someone follows a link such as those in the previous paragraph,
it's the {\em locating} form that appears in the address bar of their
browser, and is thus the form they may well copy and paste into an
email to a colleague or their own reading list.  But this undermines
the fundamental value proposition of the original ('persistent') form:
that it is not vulnerable to all the things that cause {\tt http:} URIs to
fail over time.

Our goal in the work reported here was to quantify the growth over
time in actual usage of the three forms, to see not only how good the
good news was, but also whether there was cause to worry about the
less good news: are locating forms `leaking' into public use?
 
For concrete evidence we used the Common Crawl sample of HTML pages on
the Web \cite{CC}, the only large-scale public source of evidence
readily available to us.  This turned out to be challenging in a
number of respects, to the extent that although our results are
interesting, problems with the CC data mean that they may not
accurately reflect the actual situation.  In what follows we will
first describe the work as such, and then discuss the ways in which
the CC data fell short of what we think is required for reliable
analysis.

{\bf Note on terminology}  Although most of the PIDs in various forms
(original, actionable or locating) found during our studies were DOIs,
we will be careful hereafter to use `PID' when we mean {\em anything}
recognised as a form of persistent identifier, and `DOI' for the subset
thereof which are some form of DOI.

\section{Prior work and other sources of information}

An excellent overview of the space of PIDs and arguments for their
use, only slightly dated, can be found in \cite{Duerr2011}.  The IDF's
views on the need for PIDs and their goals for DOIs is described in \cite{CCPID}.

The IDF occasionally update their "Key Facts" page
\cite{IDFkf} which currently says
that

\begin{itemize}
\item {[DOIs are]} Currently used by well over 5,000 assigners, e.g.,
  publishers, science data centres, movie studios, etc.
\item Approximately 148 million DOI names assigned to date
\item Over 5 billion DOI resolutions per year
\end{itemize}

The leading issuer of DOIs for publications is CrossRef
\cite{CrossRef}, who publish regularly-updated statistics
about membership numbers, DOIs registered,
etc. \cite{CrossRefFacts}

The leading issuer of DOIs for research data (as opposed to
publications) is DataCite \cite{DataCite} who
similarly publish statistics of the number of data-specific DOIs
issued, cited etc. \cite{DataCiteStats}

The only longitudinal study for PID usage we are aware of is \cite{Sompel2016}.  They processed
approximately 1.8 million scholarly articles published between 1997 and 2012,
drawn from arXiv.org, Elsevier journals and PubMed Central, yielding a
total of 2.2 million URIs.  Of these there were
\begin{itemize}
\item 397,412 actionable-form DOIs (all using {\tt dx.doi.org})
\item 505,657 "should-be-DOIs"
\end{itemize}

Their results are difficult to compare to ours, not only because they
were looking at a disjoint set of years, but also because they didn't
actually {\em look up} the actionable-form DOIs they found and then
tabulate the occurrences of the resulting locating-form URIs. Instead
they used "a list of hash values of publisher [domain names] provided
by CrossRef. If the hash of a [domain name] of an extracted reference
matches a hash in CrossRef’s list, a reference is [considered to be a
should-be-DOI]."  This was because their goal, as the name suggests,
was to identify references that {\em could} have been DOIs, because the
publisher was a CrossRef member and so would have assigned a DOI to
the article in question.  This is not quite the same goal as ours,
which was to measure the ratio of actionable-form to locating-form
PIDs {\em for the same individual article}.

\section{Materials}

\subsection{First study}
Our first study, of the use of PIDs in all three forms, compared
usage in April 2014 with that in April 2017, based on the Common Crawl
sample of HTML pages on the Web \cite{CC} for those months.
Table~\ref{tab:t1} gives basic size information for this sample.

\begin{table}
  \caption{Crawl size for first study \cite{Sebastian}}
  \label{tab:t1}
  \begin{tabular}{r|r|r|r}
Crawl month&URIs crawled&Pages retrieved&Dup URI \%age\\
\midrule
2014-04&1,718,646,762&2,641,371,316&34.9\%\\
\midrule
2017-04&2,907,715,349&2,942,930,482&1.2\%
\end{tabular}
\end{table}

The difference between the "URIs crawled" and "Pages retrieved"
columns in this table, particularly for 2014, signal a problem with
the same URI being retrieved multiple times.  Although the crawl
always starts with a unique set of URIs and does not follow
page-internal links, redirects to URIs in the initial set occur 
surprisingly often, giving rise to duplication in some cases.  The
"Duplicate URI \%age" column in Table~ref{tab:t1} reports this, as {\em estimated} by
subtracting the ratio of the URI to Page columns from 1. Detecting
instances of this problem and not including pages from the duplicates
has improved considerably between 2014 and 2017, as can be seen from
the convergence of the "URIs crawled" and "Pages retrieved" columns
and the big drop in the duplicate URI percentage estimate.

This duplication does not always mean that duplicate pages have been
retrieved -- as the crawl takes several weeks to complete the
identified page may have changed.  A direct estimate of the number of
duplicate pages retrieved, based on comparing Hyperloglog digest 
values, is shown in Table~\ref{tab:t2}.  We'll return to the impact this
has on our DOI tabulations in the Results section below.

\begin{table}
  \caption{Duplicate page estimates for first study \cite{Sebastian}}
  \label{tab:t2}
  \begin{tabular}{r|r|r|r}
Crawl month&Pages retrieved&Digests&Dup pages \%age\\
\midrule
2014-04&2,641,371,316&2,250,363,653&14.8\%\\
\midrule
2017-04&2,907,715,349&2,915,114,582&0.9\%
\end{tabular}
\end{table}

\subsection{Second study}
Our second study added crawls from April 2015 and 2016, but focused exclusively
on URIs using the {\tt doi:} scheme.  Table~\ref{tab:t3}
combines the columns from Tables~\ref{tab:t1}~\&~\ref{tab:t2} and includes these additional years.

\begin{table*}
  \caption{Crawl number for all four years \cite{Sebastian}}
  \label{tab:t3}
  \begin{tabular}{r|r|r|r|r|r}
Crawl month&URIs crawled&Pages retrieved&Dup URI \%age&Digests&Dup pages \%age\\
\midrule
2014-04&1,718,646,762&2,641,371,316&34.9\%&2,250,363,653&14.8\%\\
\midrule
2015-04&1,934,559,347&2,115,818,059&8.6\%&1,910,978,257&9.7\%\\
\midrule
2016-04&1,335,046,923&1,335,046,923&0.0\%&1,211,048,216&9.3\%\\
\midrule
2017-04&2,907,715,349&2,942,930,482&1.2\%&2,915,114,582&0.9\%\\
\end{tabular}
\end{table*}

The Common Crawl makes data from each crawl available in 3 variants of
the WARC format \cite{ISO28500},\cite{CCWARC}:
\begin{itemize}
  \item WARC for the raw crawl data;
  \item WAT \cite{WAT} for computed metadata, including request and response headers
    and, for responses, link tabulations from HEAD and BODY, using JSON
  \item WET for plaintext from the BODY
\end{itemize}

In both studies we worked exclusively with the WAT format, as that
contains the link data we were interested in without the additional
overhead of the entire HTML response.  The number of files, average
number of request/response pairs reported and the approximate total
compressed WAT file size (in terabytes) is shown in Table~\ref{tab:t4}.

\begin{table}
  \caption{Sizes for the second study}
  \label{tab:t4}
  \begin{tabular}{r|r|r|r}
Crawl month&WAT file count&pages per file&Total size (TB)\\
\midrule
2014-04&44488&59373&17\\
\midrule
2015-04&38609&54801&14\\
\midrule
2016-04&22200&60137&9\\
\midrule
2017-04&64700&45486&19
\end{tabular}
\end{table}

It should be noted that for 2014 and 2017, the number of actual
request/response entries recovered from the WAT files was slightly
less than the numbers published by Common Crawl: approximately 4
million less in 2014 and 600,000 less in 2017.

\section{Methods}

\subsection{First study}
For the first study we wanted to check every link from the body of
each crawled HTML page, which meant downloading around 110,000
WAT-format files totalling around 36TB in (compressed) size.

We achieved this by streaming about 1/10th of the data each night,
divided over approximately 100 machines that were detected as idle in
one of several student labs.  Each machine tabulated
summary counts for approximately 100 WAT files each night, taking
4--6 hours.  These were
uploaded to a central machine and merged.  The process was slightly
different in 2014 and 2017: only in 2017 did we look for PIDs in their
locating form, as explained below.

\subsubsection{2014 crawl}\

There were 44488 WAT files to be processed, containing information
from a total of 2,534,229,771 pages.  For each page the WAT file
contains three JSON objects, one each for information about the crawl,
the HTTP request and the HTTP response.  We extracted the latter, and
from it the following three components:

\begin{itemize}
 \item {\tt Envelope/WARC-Header-Metadata/WARC-Target-URI} (a string)
 \item {\tt Envelope/Payload-Metadata/HTTP-Response-Metadata/\
Headers/Content-Type} (a string)
 \item {\tt Envelope/Payload-Metadata/HTTP-Response-Metadata/\
HTML-Metadata/Links} (an array, see below)
\end{itemize}

For each page we accumulated counts for
\begin{itemize}
 \item the target URI scheme (always {\tt http:} or {\tt https:})
 \item the target URI host (strictly speaking the `authority' per
   RFC3986 \cite{RFC3986})
 \item the Content-Type header
\end{itemize}

The contents of the .../Links component array are each an object with
at least the following contents:
\begin{verbatim}
   { "path": [quasi-XPath, e.g. "A@/href", "IMG@/src",
              "FORM@/action"],
     "url":  [absolute or relative URI],
     [other optional properties per path]}
\end{verbatim}

For the value of the "url" property of each entry in this array we
accumulated counts for

\begin{itemize}
 \item the target URI scheme (possibly absent)
 \item the target URI host (possibly absent)
 \item if the host was one of a list of actionable PID resolvers (see
   below), the number of times the whole URI (normalised) appeared in
   the Links array
\end{itemize}

The resolvers we watched for were as follows:

\begin{verbatim}
  doi.org, dx.doi.org, dx.medra.org
  hdl.handle.net
  n2t.net
\end{verbatim}

The normalisation of the Link URIs involved
\begin{itemize}
 \item removing spurious whitespace apparently arising from issues with
   the Common Crawl process itself;
 \item replacing both percent-encoded and HTML entity-encoded character
   forms
\end{itemize}

The Links array data is our primary concern in this paper.  As the
individual processor results were merged, the individual per-page
tabulations enabled us to produce the following summary tabulation:
  
\begin{itemize}
 \item The frequency of {\tt http:} and {\tt https:} URI schemes in both the crawled URI
   set and of (none), {\tt http:}, {\tt https:} and many other URI schemes in the
   Link URI set
 \item In particular, the frequency of {\tt doi:} and {\tt info:} in the Link URI set
 \item The frequency of the five resolvers in the Link URI set
 \item The frequency of each actionable URI in the Link URI set
\end{itemize}

(Note that only a handful of actionable-form URIs appeared in the crawled URI set)

For all but the first (URI schemes in general) frequency tabulations,
we have both type and token frequency.

\subsubsection{2017 crawl}\

For the April 2017 crawl, we added two additional tabulations:

\begin{itemize}
 \item Document frequency for actionable URIs, that is, the number of
   pages in which each appears, regardless of how many times
 \item For each URI in the Link set which is the locating form of an
   actionable URI in the 2014 Link set, type, token and document
   frequency
\end{itemize}

The latter counts were tabulated by taking all the 2014 actionable
forms, issuing HTTP HEAD requests for them and noting the Location
response header which came back (iterating and accumulating until a
200 response was achieved).  This succeeded more that 99\% of the time,
and a Bloom filter was constructed from the results, which then
allowed us to check every Link URI as we processed the 2017 Link URIs.

These counts are restricted to the 2017 appearances of the locating
forms of 2014 actionable forms, because we didn't have time to do two
passes over either the 2014 data or the 2017 data.

For both years, the final step was to extract the PID itself (that is,
the path part of the actionable-form URI, regardless of URI scheme,
redirection server hostname or query parameters) and merge the counts
across all the actionable-form URIs with the same PID.  Unless
otherwise noted, these are the counts reported in the Results section
below.

\subsection{Second study}

The second study aimed to fill in the gap between 2014 and 2017, but
at a much lower level of detail.  It simply counted original-form DOI
occurrences in the HTML head (in {\btt link} and {\btt meta} elements) as well as
the body.

The scale of the 4 years' data is shown above in Table~\ref{tab:t4}.
This study was actually a pilot study to determine whether using 100
8-core computers with better bandwidth via Microsoft's Azure
facility\footnote{See \ref{sec:Ack}}
would significantly increase throughput, and did in fact allow for one
month's crawl data to be processed in about 6 hours, an improvement of
about a factor of 8 over the first study.

As in the first study, only the `response' JSON object was processed,
extracting 3 components:
  
\begin{itemize}
\item {\tt Envelope/Payload-Metadata/HTTP-Response-Metadata/\
HTML-Metadata/Links} (as in study 1)
\item {\tt  Envelope/Payload-Metadata/HTTP-Response-Metadata/\
HTML-Metadata/Head/Link} (an array)
\item {\tt  Envelope/Payload-Metadata/HTTP-Response-Metadata/\
HTML-Metadata/Head/Metas} (an array)
\end{itemize}

Each member of the Metas array is an object, where the ones of
interest had the following contents:

\begin{verbatim}
  { "name":    [the META element's name attribute]
    "content": [the META element's content attribute]}
\end{verbatim}

and we counted objects where the "content" property was an
original-form DOI.

Likewise for the Link array, where we care about

\begin{verbatim}
  { "rel":  [the LINK element's rel attribute]
    "href": [the LINK element's href attribute]}
\end{verbatim}

and counted ones where the "href" property was an original-form DOI.

In contrast to the first study, all that was tabulated were occurrence
counts per page of any original-form DOI, counts for the different DOIs
themselves were not kept, so the net results were just three totals,
first per WAT file, then after merging, per month.

Finally a very small sample, just 645 WAT files from April 2014 (1.5\%
of the total), was processed looking only at the Metas array, to count
the different values of "name" whose "content" was an original-form
DOI.

\section{Results as such}
In this section we present the results as if the data they are derived
from gave reliable evidence.  Discussion of reasons to fear this may
not be the case and suggestions for what to do about this are given in
section~\ref{sec:res2}

\subsection{First study}

Counts for all Link URIs, the actionable-form URI subset thereof and
distinct PIDs extracted from those, as tabulated across April 2014 and
2017, are shown in Table~\ref{tab:t5}.

Two columns are shown for the total number of Link URIs: The first
column is the actual number we found, the second is adjusted downwards
for the estimated degree of duplication, as reported above in
Table~\ref{tab:t1}.  This correction is not needed for the actionable
URI and PID columns (see section~\ref{sec:res2}), but is given here as
it is used for the ratios given in the Actionable Link URIs Ratio
column.

\begin{table}
  \caption{Link URI counts for first study}
  \label{tab:t5}
  \begin{tabular}{r|r|r|r|r|r}
Crawl&\multicolumn{2}{c}{Link URIs}&\multicolumn{2}{c}{Actionable Link URIs}&Distinct\\
month&Total&Corrected&URIs&Ratio&PIDs\\
\midrule
2014-04&299$ x 10^9$&194$ x 10^9$&30,445,532&0.00016&5,369,831\\
\midrule
2017-04&620$ x 10^9$&613$ x 10^9$&37,913,544&0.00006&12,659,694\\
\end{tabular}
\end{table}

The overlap between the sets of URIs crawled in April 2014 and April 2017 
is low (estimated at 7\%) and for the responses (pages) themselves even lower
(estimated at 0.8\%)
\cite{CCuniq}
However the PID numbers have a much higher overlap: the union of the
two years contains only 14.7 million PIDs -- the details are given in
Table~\ref{tab:t6}.  This suggests that the overlap PIDs are {\em very} popular, as
they are not just persisting from 2014 to 2017, but their second
appearance is in a {\em different} set of pages.

\begin{table}
  \caption{Shared vs. one-year-only PIDs in the first study}
  \label{tab:t6}
  \begin{tabular}{r|c|c}
&2014&not 2014\\
\midrule
2017&3,354,906&9,304,788\\
\midrule
not 2017&2,014,925&0\\
\end{tabular}
\end{table}

As mentioned earlier, the actionable-form PIDs that we looked for can
be divided on the basis of the domain name used to identify their
resolving proxies:  {\tt doi.org}, {\tt dx.doi.org} and {dx.medra.org}
for DOIs, {\tt hdl.handle.net} for handles and {\tt n2t.net} for ARKs
and other PIDs (we didn't explore these in any detail).  The numbers
in each category are shown in Table~\ref{tab:t6a}.

\begin{table}
  \caption{PID scheme  and resolver counts from distinct actionable-form PIDs}
  \label{tab:t6a}
  \begin{tabular}{r|r|r|r}
When&DOIs&handles&other\\
\midrule
2014 only&1,656,913&357,997&15\\
\midrule
2014 and 2017&2,914,930&439,969&7\\
\midrule
2017 only&7,383,189&1,914,395&7,204
\end{tabular}
\end{table}

The relative recency of the arrival of the {\tt n2t.net} resolver on
the scene is clearly evident here.

Finally the locating-form leakage question is addressed in
Table~\ref{tab:t7}, which gives the number of locating-form URIs retrieved for the
actionable-form URIs found in 2014.

\begin{table}
  \caption{Locating-form URIs in 2017 for 2014 actionable-form PIDs}
  \label{tab:t7}
  \begin{tabular}{r|r|r}
&Distinct&Total\\
\midrule
Actionable found in 2014&5,369,831&12,642,054\\
\midrule
Retrieved locating form&5,315,129&\\
\midrule
Locating found in 2017&413,397&1,202,610\\
\midrule
Ratio&8\%&10\%
\end{tabular}
\end{table}

There were 12+ million actionable-form URIs
found in 2014, from which 5+ million distinct PIDs were extracted,
almost all of which successfully yielded locating URIs.  Of
these around 400,000 (8\%) by type count, or 1.2 million (10\%) by token count, occurred in body links
in the the 2017 crawl.  There is of course no way to tell whether
these usages arose from the kind of leakage scenario discussed in the
Introduction, or whether they were found and used independently of the
antecedent actionable-form URI, but either way this is a large enough
number to be of some concern.

\subsection{Second study}

Adding data for April 2015 and 2016 allows us to track the growth of
{\tt doi:} use in HTML body and head links (distinguishing between
{\btt meta} and {\btt link} elements).  In head {\btt link} elements
we found no (!) uses of the {\tt doi:} form in April 2014 or 2015, and
only 2 in April 2016 and 2017, so the data in Table~\ref{tab:t8}, with
graphs in Figure~\ref{fig:f1}, only report on the
numbers for use in body links and head {\btt meta} elements.

\begin{table}
  \caption{Growth of {\tt doi:} usage for second study}
  \label{tab:t8}
  \begin{tabular}{r|r|r|r|r}
\multicolumn{2}{c}{Body links}&\multicolumn{2}{c}{Head {\btt meta}}\\
year&n&per mil pg&n&per 10K pg\\
\midrule
2014&1893&0.72&731938&2.77\\
\midrule
2015&1410&0.67&727167&3.44\\
\midrule
2016&1440&1.08&410603&3.08\\
\midrule
2017&3550&1.21&459328&1.56
\end{tabular}
\end{table}

\label{fig:f1}
\begin{figure}
  \caption{Growth in {\tt doi:} use in body links and in head meta}
  \centering
    \includegraphics[width=0.50\textwidth]{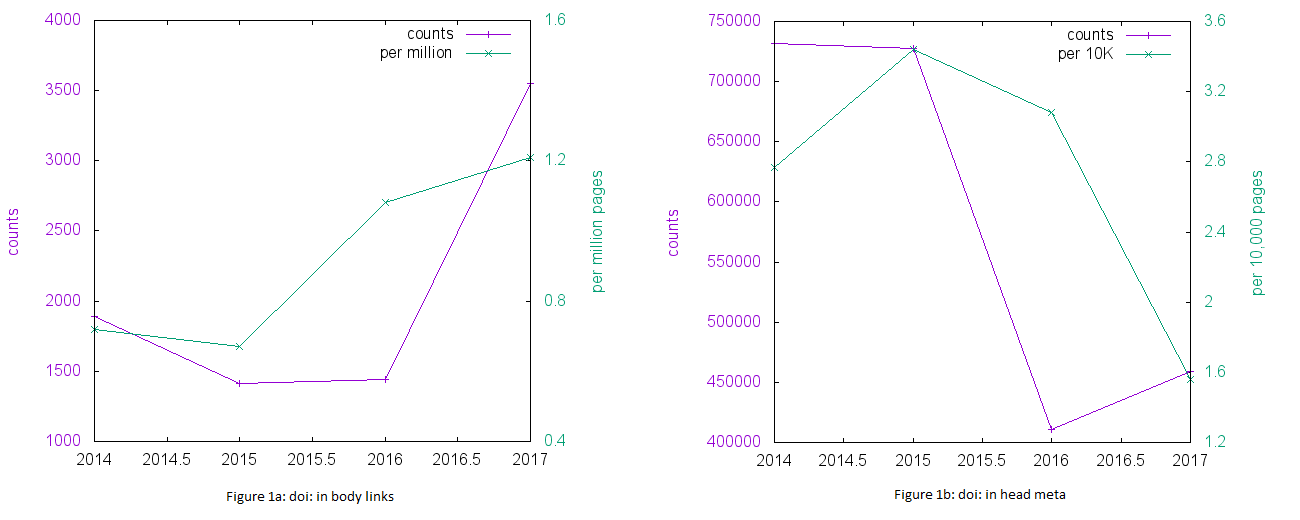}
\end{figure}


It's interesting to see that a small number of original-form {\tt
  doi:} URIs are appearing as e.g. {\btt A/@href}, and that this usage
is slowly increasing.  It's certainly not obvious {\em why} anyone
would do this:  It would be necessary to look at the complete HTML
pages to make sense of this.

The much more substantial use in HTML head {\btt meta} elements is,
on the other hand, quite plausible, although the drop in 2017 is hard
to evaluate without seeing data from the surrounding months.

We did a quick check of around 1.5\% of the April 2014 to see which
meta tags the doi: URIs were being used with.  Table~\ref{tab:t9} gives the
rank-ordered results.

\begin{table}
  \caption{What {\tt meta} tags are {\btt doi:} URIs used for?}
  \label{tab:t9}
  \begin{tabular}{r|r||r|r}
Tag&Count&Tag&Count\\
\midrule
dc.identifier&6548&dcterms.isReferencedBy&4\\     
\midrule
eprints.id\_number&1174&eprints.related\_url\_url&2\\  
\midrule
citation\_doi&435&keywords&2\\                   
\midrule
dc.Identifier&146&bepress\_citation\_doi&1\\     
\midrule
dcterms.isVersionOf&105&dc.citation.spage&1\\          
\midrule
dc.relation&44&eprints.data&1\\               
\midrule
dcterms.hasPart&15&eprints.doi&1\\                
\midrule
dcterms.isPartOf&12&eprints.note&1\\               
\midrule
&&eprints.official\_url&1
\end{tabular}
\end{table}

The vast majority of these are Dublin Core \cite{DublinCore} or EPrints
toolset \cite{EPrints}
tags.

Given the quite large number of doi: URIs showing up in the HTML
header as {\btt META/@content}, it's a bit surprising not to
find any as {\btt LINK/@uri}.

\subsection{Conclusions for DOIs}

In summary, the results of the two studies show
\begin{itemize}
 \item Virtually no use of original-form URIs in head links
 \item Only small numbers of 1000s of original-form URIs in body links
 \item Significant, slowly increasing, use (100s of thousands) of original-form URIs as meta-information
 \item Much larger numbers (millions) of actionable-form URIs in body links
 \item A 2.5 times increase in the number of distinct DOIs in body
   links between the 2014 and 2017 crawls
 \item For about 8\% of the actionable-form URIs in the 2014 crawl,
   the corresponding locating-form URI appears in the 2017 crawl
\end{itemize}

\section{Conclusions for longitudinal studies}\label{sec:res2}
The work reported here cannot be taken as anything more than a starting point,
demonstrating that it is possible to extract longitudinal
information about URI usage and encouraging others to do so: neither the
numbers nor the trends presented here can be claimed to be reliable.
It illustrates the kind of questions we would like to get answers for
with respect to one kind of longitudinal study, and the very {\em
lack} of reliable answers to those questions points towards the things
we need to do to improve the situation.

In what follows we look at a number of different kinds of problem we
encountered and suggest possible remediations.

\subsection{Common Crawl itself}
Duplication of pages crawled within an individual release is an issue
for any use of Common Crawl data.  Duplication of pages between
releases may be a bug or a feature for longitudinal study, but its
existence needs to be taken account of in any case.

A number of discussions of these issues can be found in the Common
Crawl forum \cite{ccf}, and it does appear that within-release
duplication has been considerably reduced.

However, for our study, and as noted above in the
Materials section, the April 2014 data shows a substantial degree of
likely duplication at the page level.  This is almost entirely due to
two sources \cite{Tom}
\begin{enumerate}
 \item Shared error pages;
 \item Same page for distinct URIs differing only in query parameter values.
\end{enumerate}

It seems reasonable to assume that error pages are unlikely to involve
much PID use, and the same is true for the kinds of
commercially-orientated applications which make heavy use of query
parameters.  The latter expectation is easy to check empirically, and
a quick check of a random sample (3425 actionable-form (a mixture of
doi.org and dx.doi.org) URIs from 4 different WAT files from April
2014) confirms this: none of them have query parameters.

There's clearly a pressing need for a careful study of the last 3 or 4
years of CC data, to establish in detail the within- and
between-release overlaps, both with respect to content and URI (see
also section~\ref{sec:vers}).  CC's own version of this information
\cite{CCuniq} covers 2015 onwards, but has not as far as we know been
published in a peer-reviewed context or otherwise confirmed.

In harvesting URIs from the Links component of the response records in
a CC WAT file, we encountered a wide range of low-level problems with
format and character encoding.  Some of these did {\em not} occur in
the original, in the few cases we checked by hand and could find.
Although tedious, a survey of the kinds of errors introduced in the
WAT files is needed, to at least document their frequency over time,
but also to try to establish which can be detected reliably and, of
those detectable, which can be reliably corrected.  For those problems
which persist in more recent releases, we would hope once alerted to
them CC could fix the problem at source going forward.

Correct reporting of links that are found is important, but so is
actually reliably {\em detecting} links---some empirical checking of
this would also be a good idea.

The recent rapid growth of personalised responses based on information
in the query string of URIs and/or on cookies has serious implications
for the 'Identifier' aspect of URIs, and the extent to which responses
to URIs which share their {\tt authority} and {\tt path} components
but not their {\tt query}.  Again, at least some comparison of CC
crawl target URIs both with and without including the {\tt query}
component is needed.

For the two CC releases we have checked, that is April 2014 and April
2017, the good news is that the percentage of HTTP vs. HTTPS is close
as between URIs crawled and Link URIs seen (around 20 to 1 in 2014,
dropping to 3.3 to 1 in 2017, reflecting the success of initiatives
such as Encrypt the Web \cite{Lets}.  A more systematic tabulation over
all releases is obviously needed before we can tell whether this is a
reliable trend or not.  How representative a sample the CC HTML is of
Web HTML as a whole is unknown, and indeed it's not clear how one
would quantify this.

A {\em much} more serious coverage issue, particularly for the persistence
issues we started out to explore, is the lack of anything other than
HTML documents in the CC releases.  For scholarly publications, which
are a major market for PIDs, PDF is the preferred format for
publication.  Expanding CC to include PDF files clearly would be a
major undertaking, but at least some attempt to crawl links from a CC
release to PDF files would be useful to get some sense of how much the
profile of links found {\em there} differs from that in the HTML data.

\subsection{Versioning and deduplication}\label{sec:vers}
Detecting and at least tabulating, preferably eliminating, exact
duplicate content is of course important, but for at least some kinds
of longitudinal studies, detecting and relating multiple versions of
the 'same' content is also important.  Detecting similar-but-different
content retrieved from different (post-redirection) URIs is obviously
non-trivial, as it depends implicitly on some notion of 'sufficient'
similarity to count as the same version.  Plagiarism detection
software has a contribution to make here.  Even quite a tight
threshold might be very useful: in a way two documents which differ
only by a tiny change, say a single spelling correction, is much worse
than two identical document, because hash-based methods will find the
latter but not the former.

\subsection{Scale}

At the very least the variability and occasional unreliability of the
Common Crawl data means that for improved confidence the usage being
studied should be tabulated for {\em every} month's crawl over a at
least a year.  With hindsight it would also probably be wise to use
crawls from 2015 onwards, as both the effort to remove duplicates and
the documentation improve noticeably at that point.  This in turn
however begins to move the effort involved out of the reach of the
kind of {\em ad-hoc} multiprocessor we assembled and used for the
first study.  Even the 6-hour turnaround we achieved for the second
study still made debugging a tedious and potentially expensive
process.  We have some speed-ups in mind, but they are unlikely to
gain us more than a factor of two or so.  Generosity of the sort
provided by the donor of cloud resources for the second study will be
needed if academic longitudinal studies of Web usage are to reach the
levels of reliability and utility we need.

In conclusion, Common Crawl releases since 2015 provide a potential
basis for longitudinal studies of HTML web page content and linking,
but results have to be treated with caution.  A number of gaps in
documentation and quality assurance need to be addressed before
conclusions based on such studies can be taken as reliable.

\label{sec:Ack}\begin{acks}
  The idea for this kind of study came in the form of a question by
  Greg Jan\'{e}e at PHOIBOS 2 \cite{PHOIBOS2}: ``How effective are
  PIDs?''.  He subsequently expanded on this: ``A PID is {\em effective} only if the
  resource identified by the PID is uniformly and universally
  accessed via the PID, and not via other non-persistent URLs'' \cite{GJ}.
  
Thanks to Sebastian Nagel of Common Crawl for prompt and helpful
answers to my questions about duplicates and duplicate detection.

The second study reported above was made possible by Microsoft's
donation of Azure credits to The Alan Turing Institute. This work
was supported by The Alan Turing Institute under the EPSRC grant
EP/N510129/1.  Our thanks to Dr. Kenneth Heafield (Informatics,
University of Edinburgh) for his help with using this.
\end{acks}